\documentclass[twocolumn,prl,letter,showpacs,floatfix]{revtex4}

\usepackage{graphicx}

\begin{document}

\title{Superconductivity and Quantum Criticality in CeCoIn$_5$ }

\author{V.~A.~Sidorov}
 \altaffiliation{Permanent address: Institute for High Pressure
 Physics, Russian Academy of Sciences, Troitsk, Russia.}
\author{M.~Nicklas}
\author{P.~G.~Pagliuso}
\author{J.~L.~Sarrao}
\author{Y.~Bang}
\author{A.~V.~Balatsky}
\author{J.~D.~Thompson}

\affiliation {Los Alamos National Laboratory, Los Alamos, NM
87545}

\date{\today}

\begin{abstract}
Electrical resistivity measurements on a single
crystal of the heavy-fermion superconductor CeCoIn$_5$ at
pressures to 4.2~GPa reveal a strong crossover in transport
properties near $P^*\approx1.6$~GPa, where $T_c$ is a maximum.
The temperature-pressure phase diagram constructed from these
data provides a natural connection to cuprate
physics, including the possible existence of a pseudogap.
\end{abstract}


\maketitle

The relationship between unconventional superconductivity and
quantum criticality is emerging as an important issue in strongly
correlated materials, including cuprates, organics and
heavy-fermion intermetallics \cite{orenstein00,chubukov02}. In
each of these, either Nature or an externally imposed parameter,
such as pressure or chemical substitutions, tunes some magnetic
order to a $T=0$ transition where quantum fluctuations introduce
new excitations that control thermodynamic and transport
properties over a broad range of temperature-parameter phase
space. A frequent, but controversial \cite{anderson02},
assumption is that these same excitations may mediate Cooper
pairing in the anisotropic superconductivity that appears in the
vicinity of a quantum-critical point (QCP). Indeed, the effective
potential generated by proximity to a quantum-critical
spin-density wave \cite{coleman01a} also can lead
\cite{montoux91} to anisotropic superconductivity.

Though the issue of quantum criticality and superconductivity
came to prominence in the context of the cuprates (see, eg
\cite{varma01}), there have been several recent examples of this
interplay in heavy-fermion compounds \cite{mathur98}. Like the
cuprates, unconventional superconductivity in these Ce-based
heavy-fermion materials develops out of a distinctly
non-Fermi-liquid normal state that evolves in proximity to a
continuous $T=0$ antiferromagnetic transition; but unlike the
cuprates, this state can be accessed cleanly by applied
hydrostatic pressure and without introducing extrinsic disorder
associated with chemical substitutions. In addition to chemical
inhomogeneity in the cuprates, a further 'complication' is the
existence of a pseudogap state above $T_c$ for dopings less than
optimal \cite{timusk99}.  Whether evolution of the pseudogap with
doping produces another QCP near the optimal $T_c$ in the cuprates
is an open question as is the origin of the pseudogap
\cite{varma99}. Thus far, a pseudogap has not been detected in
canonical quantum-critical heavy-fermion systems, such as
CeIn$_3$ or CePd$_2$Si$_2$ under pressure, but this may not be
too surprising. In these cases, superconductivity appears only at
very low temperatures, less than 0.4~K, and over a narrow window
of pressures $(\leq0.8 {\rm~GPa})$ centered around the pressure
where $T_N$ extrapolates to zero. In analogy to the cuprates, we
would expect a pseudogap in these Ce compounds to exist only over
a narrow temperature window, $0.1-0.2$~K above $T_c$, and in a
fraction of the narrow pressure window, which certainly would
make detection of a pseudogap difficult.

CeCoIn$_5$ offers the possibility of making a connection between
other Ce-based heavy-fermion materials and the cuprates. Bulk,
unconventional superconductivity is present in CeCoIn$_5$ at
atmospheric pressure
\cite{petrovic01b,movshovich01,izawa01,kohori01} and develops out
of a heavy-fermion normal state in which the resistivity is
approximately linear in temperature. Application of a magnetic
field sufficient to quench superconductivity reveals a specific
heat diverging as $-T\ln T$ and associated low-temperature
entropy consistent with its huge zero-field specific heat jump at
$T_c$ $(\Delta C/\gamma T_c=4.5)$. Together, these properties
suggest that CeCoIn$_5$ may be near an antiferromagnetic
quantum-critical point at $P=0$. Its isostructural,
antiferromagnetic relative CeRhIn$_5$ becomes a pressure-induced
unconventional superconductor near 1.6~GPa, with a $T_c=2.1$~K
close to that of CeCoIn$_5$ at $P=0$ \cite{hegger00,fisher01}.
This and other similarities between CeCoIn$_5$ at $P=0$ and
CeRhIn$_5$ at $P=1.6$~GPa reinforce speculation
\cite{nicklas01,shishido02} that the nearby antiferromagnetic QCP
in CeCoIn$_5$ may be at an inaccessible slightly negative
pressure. The 1.7\% smaller cell volume of CeCoIn$_5$ compared to
that of CeRhIn$_5$ is consistent with this view. We have studied
the effect of pressure on the electrical resistivity and
superconductivity of CeCoIn$_5$ and find striking correlations
between them that are reminiscent of cuprate behaviors.

Four-probe AC resistivity measurements, with current flowing in
the tetragonal basal plane, were made on a single crystal of
CeCoIn$_5$ grown from excess In flux. Hydrostatic pressures to
5~GPa were generated in a toroidal anvil cell
\cite{khvostantsev98} in which a boron-epoxy gasket surrounds a
teflon capsule filled with a glycerol-water mixture (3:2 volume
ratio) that served as the pressure transmitting fluid. Pressure
inside the capsule was determined at room temperature and at low
temperatures from the pressure-dependent electrical resistivity
and $T_c$ of Pb, respectively \cite{eiling81}. The width of
superconducting transition of Pb did not exceed 15~mK, indicating
good hydrostatic conditions and providing an estimate of the
pressure-measurement uncertainty, $\pm0.04$~GPa.

\begin{figure}[t]
\includegraphics[angle=0,width=80mm,clip]{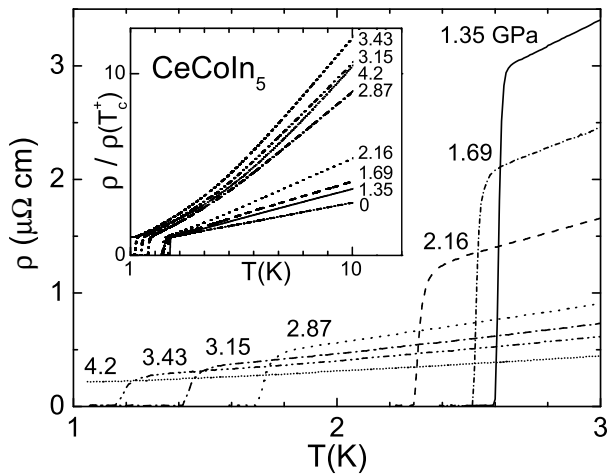}
\caption{\label{fig1} Effect of pressure on the low-temperature
resistivity and superconducting $T_c$ of CeCoIn$_5$. The inset
shows the resistivity $\rho$ normalized by $\rho(T^{+}_{c})$ up to
10~K.}
\end{figure}

\begin{figure}[t]
\includegraphics[angle=0,width=80mm,clip]{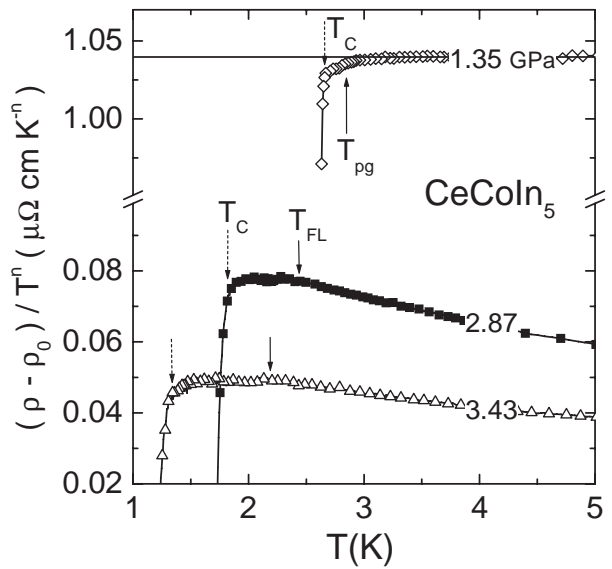}
\caption{\label{fig2} Representative plots of
$(\rho(T)-\rho_0)/T^n$ vs $T$. At low pressures,
$(\rho(T)-\rho_0)/T^n$ is constant from $T_{pg} \approx
(1.15\pm0.05) T_c$ to about 10 K. At the higher pressures, there
is a well-defined range below $T_{FL}$ where $(\rho-\rho_0)\propto
T^2$ is Fermi-liquid like. The resistive onset of
superconductivity, defined by the intersection of linear
extrapolations of $\rho(T)$ from above and below $T_c$, is denoted
by dashed vertical arrows. At the highest pressures, the gradual
rounding just above $T_c$ is due to a broadened transition (see
Fig.~\ref{fig3}) and not to $T_{pg}$}
\end{figure}

The response of $\rho(T)$ to pressure over a broad temperature
scale \cite{nicklas01,shishido02} is typical of many Ce-based
heavy-fermion compounds and reflects a systematic increase in the
energy scale of spin fluctuations. Most interesting is the
low-temperature response (Fig.~\ref{fig1}) where $T_c$ and the
resistivity just above $T_c$ $(\rho(T_c^+))$ decrease rapidly
with pressures $P>1.35$~GPa. We have analyzed the low temperature
resistivity, plotted over a broader temperature scale in the inset
of Fig.~\ref{fig1}, by either fitting to a form $\rho(T) =
\rho_0+AT^n$, with $\rho_0$, $A$ and $n$ fitting parameters, or
from plots of $(\rho(T) - \rho_0)/T^n$. A drawback to the former
approach is that parameters can be sensitive to the temperature
range chosen for fitting. The more sensitive and revealing
approach is the latter in which $\rho_0$ and $n$ are adjusted to
give the best horizontal line. This procedure requires one less
parameter, $A$, which is provided directly by the magnitude of
$(\rho(T) - \rho_0)/T^n$. Representative examples of this
approach are plotted in Fig.~\ref{fig2}. Parameters obtained from
these plots are well defined and agree, typically to 10\% or
better, with parameter values extracted from straightforward fits.
As indicated by arrows on the lowest pressure curve in
Fig.~\ref{fig2}, $(\rho(T) - \rho_0)/T^n$ deviates from the
horizontal trend at a temperature $T_{pg}$ well above the onset
of superconductivity. This departure implies either a decrease in
the scattering rate or increase in carrier density. Resistivity
measurements at $P=0$ and in a magnetic field (6~T) greater than
$H_{c2}(0)$ find \cite{movshovich02} a similar departure of
$(\rho(T) - \rho_0)/T^n$ from a $T$-independent curve at
$T_{pg}$=3 K. Thus, $T_{pg}$ does not originate from
superconducting fluctuations or from trace amounts of free
indium. We also note that this behavior was present in previous
measurements on other single crystals CeCoIn$_5$ but was missed
in inspections of $\rho$ versus $T$ curves \cite{nicklas01}. We
will return to a discussion of $T_{pg}$ later.

\begin{figure}[t]
\includegraphics[angle=0,width=80mm,clip]{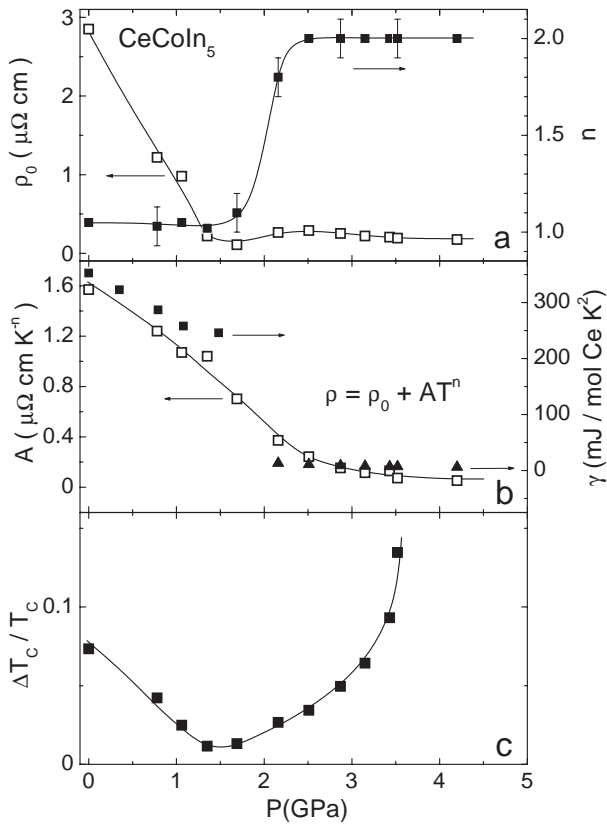}
\caption{\label{fig3} (a) Values of $\rho_0$ and $n$ obtained from
plots as shown in Fig.~\ref{fig2}. (b) Temperature coefficient of
resistivity, $A$, associated with values of n given in the top
panel, and specific heat coefficient $\gamma$ measured directly
($P < 1.5$ GPa) and,  for $P> 2$ GPa , inferred from the $T^2$
coefficient of $\rho(T)$. See text for details. (c)
Superconducting transition width normalized by $T_c$. $\Delta
T_c$ and $T_c$ are the half-width of the resistive transition
width and resistive midpoint, respectively. Solid lines in all
cases are guides to the eye.}
\end{figure}

Values of $\rho_0$, $n$ and $A$ obtained from these plots are
summarized in Figs.~\ref{fig3}a and b. There is an unmistakable
crossover in the pressure dependence of $\rho_0$ and $n$ at
$P^*\approx1.6$~GPa. Below $P^*$, $n=1.0\pm 0.1$ for $T_{pg}\leq
T \leq10$~K. This value of $n$ is expected for a 2-dimensional,
antiferromagnetic quantum-critical system \cite{sachdev99} and is
not too surprising given the layered crystal structure
\cite{moshopoulou01} and anisotropy in electronic states of
CeCoIn$_5$ \cite{petrovic01b,settai01}. At pressures greater than
$P^*$, $n$ rapidly approaches the Fermi-liquid value of 2.0,
which holds from just above $T_c$ to $T_{FL}$ (see Fig.~2).
Though not shown in either Fig.~\ref{fig2} or \ref{fig3}a, at
these higher pressures $n$ assumes a value of $1.5\pm 0.1$,
characteristic of a 3-D antiferromagnetic QCP, in a temperature
interval from slightly greater than $T_{FL}$ to as high as
$\sim$\,60~K. For $P\geq P^*$, $\rho_0$ decreases reversibly by
an order of magnitude to a very small value of about
0.2~$\mu\Omega{\rm cm}$. Clearly, pressure does not remove
impurities from the sample; the large decrease in $\rho_0$ must
be due to a pressure-induced change in inelastic scattering
processes. Theories of electronic transport in quantum-critical
systems show \cite{rosch99,miyake01} that impurity scattering can
be strongly enhanced by quantum-critical fluctuations. This
sensitivity provides a natural explanation for a decrease in
scattering with increasing pressure, if there is a crossover near
$P^*$ from quantum-critical $(P<P^*)$ to a Fermi-liquid-like
state for $(P>P^*)$. Such a crossover is suggested by the
pressure variation of $n$.

Experimental values of the $T^n$-coefficient of resistivity,
plotted in Fig.~\ref{fig3}b, also decrease strongly with
increasing pressure for $P \lesssim P^*$. For comparison, we show
the variation of the electronic specific heat divided by
temperature $(C/T\equiv \gamma)$ at 3~K for pressures less than
1.5~GPa \cite{sparn01} and, at higher pressures, we plot values
of $\gamma$ inferred from the empirical relationship
$A=1\times10^{-5}\gamma^2$, where $A$ is the $T^2$-coefficient of
resistivity, followed by several heavy-fermion compounds
\cite{kadowaki86}. Though a $T^2$ coefficient proportional to
$\gamma^2$ is expected for a Landau Fermi-liquid \cite{varma01},
it is not obvious {\it a priori} why $A(P)$ approximately tracks
the directly measured $\gamma$ at low pressures, but this seems
to be the case. In any event, both $A(P)$ and $\gamma(P)$ exhibit
qualitatively different magnitudes and functional dependencies
above and below $P^*$. The dramatic differences in $\gamma(P)$ at
low and high pressures suggest that either the empirical
relationship is invalid and underestimates $\gamma$ in the
high-pressure regime by at least two orders of magnitude or there
is a very rapid crossover in the density of low-energy
excitations in the vicinity of $P^*$. In the Fermi-liquid-like
state above $P^*$, we would expect the temperature scale $T_{FL}$
(see Fig. 2) to be proportional to $A^{-0.5}$ instead of the
approximate $T_{FL}\propto \sqrt{A}$ found experimentally. We do
not understand this discrepancy. Direct measurements of $C/T$ at
pressures above $P^*$ would be valuable.

\begin{figure}[t]
\includegraphics[angle=0,width=80mm,clip]{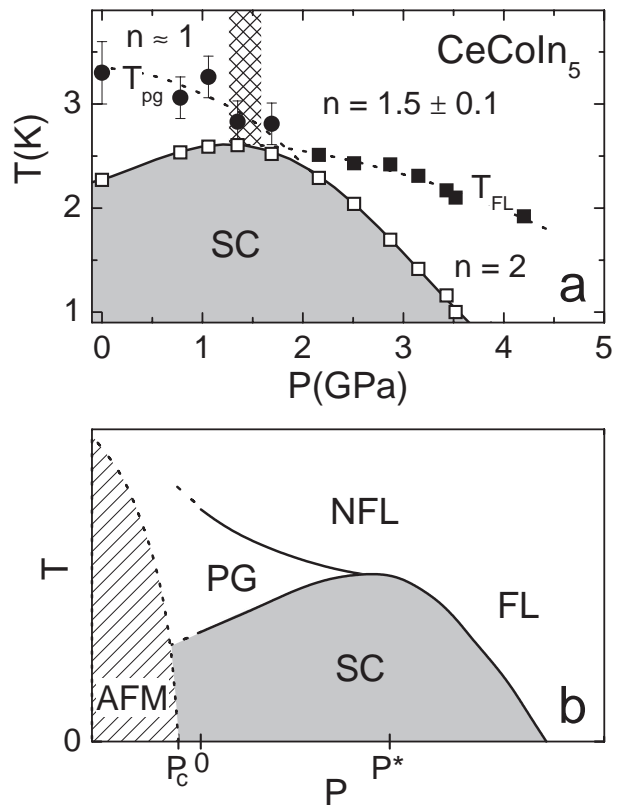}
\caption{\label{fig4} (a) Temperature-pressure phase diagram for
CeCoIn$_5$ constructed from data shown in Figs.~\ref{fig2} and
\ref{fig3}. (b) Schematic $T{\rm-}P$ phase diagram. AFM: N\'{e}el
state; PG: pseudogap state; SC: unconventional superconducting
state; FL: Fermi liquid; NFL: non-Fermi-liquid. See text for
details.}
\end{figure}

Finally, we note that the resistive transition to the
superconducting state is equally sensitive to the crossover at
$P^*$. As shown in Fig.~\ref{fig3}c, the relative transition
width $\Delta T_c/T_c$ passes through a pronounced minimum near
$P^*$. To some extent this behavior is accentuated by the pressure
variation of $T_c$, but $\Delta T_c$ itself is a minimum near
$P^*$. It is as if the optimally homogeneous state is singular in
the vicinity of $P^*$.

From the data presented above, we construct a
temperature-pressure phase diagram for CeCoIn$_5$ given in
Fig.~\ref{fig4}a. This phase diagram is more like that of the
cuprates around optimal doping than the canonical behavior of
heavy-fermion systems exemplified, for example, by CeIn$_3$ or
CePd$_2$Si$_2$ \cite{mathur98}. There is no obvious long-range
ordered state in CeCoIn$_5$ at atmospheric or higher pressures.
Rather, in view of its relationship to CeRhIn$_5$, a reasonable
speculation is that the antiferromagnetic QCP of CeCoIn$_5$ is at
a slightly negative pressure. The resistivity exponent $n$ and
specific heat (ref. \cite{sparn01}) are those of a
non-Fermi-liquid up to $P^*$ where $T_c$ is a maximum. Beyond
$P^*$, $T_c$ drops rapidly and parameters characterizing
electronic transport are qualitatively different,
Fermi-liquid-like at temperatures just above $T_c$ and those of a
3-D quantum-critical antiferromagnetic state at higher
temperatures.

The generic $T{\rm-}P$ phase diagram in Fig.~\ref{fig4}b  provides
a broader perspective of our experimental observations and their
relationship to CeRhIn$_5$. In the spirit of this diagram, the
non-Fermi-liquid transport and thermodynamic properties of
CeCoIn$_5$ below $P^*$ are controlled by an antiferromagnetic QCP
at the inaccessible negative pressure denoted by $P_c$ in
Fig.~\ref{fig4}. We have chosen $P_c$ to correspond approximately
to the critical pressure at which superconductivity appears in
CeRhIn$_5$. Spin-lattice relaxation measurements on CeRhIn$_5$
identify \cite{kawasaki01} a pseudogap whose signature first
appears at pressures slightly less than $P_c$ and at a
temperature of $\sim$5~K and then decreases toward $T_c$ with
increasing pressure until its signature is lost at $P\gtrsim
P_c$. If we associate the temperature $T_{pg}(P)$ in
Fig.~\ref{fig4}a with a resistive signature for a pseudogap and
extrapolate $T_{pg}(P)$ in CeCoIn$_5$ to slightly negative
pressures, there is a natural connection between our observations
and properties of CeRhIn$_5$. Neither the $1/T_1T$ measurements
on CeRhIn$_5$ nor our transport measurements on CeCoIn$_5$
permits definitive statements about the origin or nature of the
pseudogap \cite{pseudogap}. We note, however, that 4-fold
anisotropy of the thermal conductivity in the basal plane of
CeCoIn$_5$ persists \cite{izawa01} to $3.2{\rm~K}\approx T_{pg}$
at atmospheric pressure. If this in-plane modulation of the
normal-state thermal conductivity is due to the presence of a
pseudogap, it implies d-wave symmetry.

The schematic phase diagram, Fig.~\ref{fig4}b, reflects the
relationship among various phases of CeCoIn$_5$ and CeRhIn$_5$
and is similar to the $T$-doping phase diagram of the cuprates.
This diagram and experimental data from which it is inferred
indicate that the physics of heavy-fermion systems may be more
closely related to that of the cuprates than previously
appreciated. Aside from the structural layering in CeCoIn$_5$ and
CeRhIn$_5$ and their relatively high $T_c$'s, they are similar in
many respects to other Ce-based heavy-fermion materials in which
superconductivity develops near a QCP. Further study of
CeCoIn$_5$, in parallel with the cuprates and other
pressure-induced heavy-fermion superconductors, holds promise for
bridging our understanding of the inter-relationship between
unconventional superconductivity and quantum criticality across
these interesting classes of correlated matter.

We thank R. Movshovich for helpful discussions. VS acknowledges
S.~M. Stishov for his support during the initial stage of this
research. Work at Los Alamos was performed under the auspices of
the US Department of Energy.

\newpage

\vspace{10cm}


\begin{thebibliography}{100}

\bibitem{orenstein00}J. Orenstein and A. J. Millis, Science {\bf
288}, 468 (2000); S. Sachdev, {\it ibid.} {\bf 288}, 475 (2000).

\bibitem{chubukov02} A. Y. Chubukov, D. Pines, and J. Schmalian,
cond-mat/0201140.

\bibitem{anderson02}P. W. Anderson, Physica B {\bf 318}, 28 (2002).

\bibitem{coleman01a}P. Coleman and C. Pepin, Physica B {\bf 312-313}, 383 (2002).

\bibitem{montoux91}P. Monthoux, A. V. Balatsky, and D. Pines,  Phys. Rev. Lett. {\bf67}, 3448 (1991).

\bibitem{varma01}C. M. Varma, Z. Nussinov, and W. van Saarloos, Phys. Rep. {\bf 361}, 267 (2002).

\bibitem{mathur98} N. D. Mathur {\it et al.}, Nature {\bf 394}, 39
(1998).

\bibitem{timusk99} T. Timusk and B. Statt, Rep. Prog. Phys. {\bf62}, 61 (1999).

\bibitem{varma99} See, for example, C. M. Varma, Phys. Rev. Lett. {\bf83}, 3538 (1999);
 A. Sokol and D. Pines, Phys. Rev. Lett. {\bf71}, 2813 (1993);
 S. Chakravarty {\it et al.}, Phys. Rev. B {\bf63}, 094503 (2001);
 Ar. Abanov, A.V. Chubukov, and J. Schmalian, Europhys. Lett. {\bf55}, 369 (2001); and refernces
therein.

\bibitem{petrovic01b}C. Petrovic {\it et al.}, J. Phys.: Condens. Matter {\bf 13}, L337
(2001).

\bibitem{movshovich01}R. Movshovich {\it et al.}, Phys. Rev. Lett. {\bf 86}, 5152
(2001).

\bibitem{izawa01}K. Izawa {\it et al.},Phys. Rev. Lett. {\bf87}, 057002
(2001).

\bibitem{kohori01}Y. Kohori {\it et al.}, Phys. Rev. B {\bf 64},
134526 (2001).

\bibitem{hegger00}H. Hegger {\it et al.}, Phys. Rev. Lett. {\bf 84},
4986 (2000).

\bibitem{fisher01}R. A. Fisher {\it et al.}, Phys. Rev. B {\bf 65}, 224509 (2002).

\bibitem{nicklas01}M. Nicklas {\it et al.}, J. Phys.: Condens. Matter {\bf
13}, L905 (2001).

\bibitem{shishido02}H. Shishido {\it et al.},  J. Phys. Soc. Jpn. {\bf 71}, 162 (2002).


\bibitem{khvostantsev98} L. G. Khvostantsev, V. A. Sidorov, and O. B. Tsiok, in: {\it Properties
of Earth and Planetary Materials at High Pressures and
Temperatures}, Ed. by M.H. Manghnani and T. Yagi, Geophysical
Monograph {\bf101}, American Geophysical Union, 1998, p.89.

\bibitem{eiling81} A. Eiling and J. S. Schilling, J. Phys. F: Metal Phys. {\bf11},
623 (1981).

\bibitem{movshovich02}R. Movshovich, private communication.

\bibitem{sachdev99} S. Sachdev, {\it Quantum Phase Transitions} (Cambridge University Press, Cambridge, 1999).

\bibitem{moshopoulou01}E. G. Moshopoulou {\it et al.}, J. Solid State Chem.
{\bf 158}, 25 (2001).

\bibitem{settai01}R. Settai {\it et al.}, J. Phys.: Condens. Matter {\bf
13},
L627 (2001).

\bibitem{rosch99}A. Rosch, Phys. Rev. Lett. {\bf82}, 4280 (1999).

\bibitem{miyake01} K. Miyake and O. Narikiyo, J. Phys. Soc. Jpn. {\bf 71}, 867 (2002).

\bibitem{sparn01}G. Sparn, {\it et al.} Physica B {\bf 312-313}, 138 (2002);
E. Lengyel {\it et al.} High Pressure Res. {\bf 22}, 185 (2002).
In these references, for $H\geq H_{c2}(0)$ $C/T=-\ln T$ persists
to $P=1.48$~GPa, but the rate of divergence decreases with
increasing $P$.

\bibitem{kadowaki86}K. Kadowaki and S. B. Woods, Solid State Comm.
{\bf 58}, 507 (1986).

\bibitem{kawasaki01}S. Kawasaki {\it et al.}, Phys. Rev. B {\bf 65}, 020504
(2001).

\bibitem{pseudogap}There is no obvious evidence for a pseudogap at $P=0$ in
specific heat or magnetic susceptibility data for CeCoIn$_5$,
possibly because the signature is expected to be weak and the
temperature range between $T_{pg}$ and $T_c$ is limited to no
more than 1.1~K.

\end{thebibliography}
\end{document}